\documentclass[aps,groupedaddress,showpacs,showkeys,nofootinbib]{revtex4}
\usepackage{graphicx,amssymb,amsmath,bm}

\newcommand{\be}{\begin{equation}}
\newcommand{\ee}{\end{equation}}
\newcommand{\bea}{\begin{eqnarray}}
\newcommand{\eea}{\end{eqnarray}}
\newcommand{\nn}{\nonumber}

\def\de#1/de#2{\frac{\partial {#1}}{\partial {#2}}}
\begin{document}
\title{Critical exact solutions for self-gravitating Dirac fields}
\author{Roberto Cianci\footnote{E-mail: cianci@dime.unige.it},
Luca Fabbri\footnote{E-mail: fabbri@dime.unige.it},
Stefano Vignolo\footnote{E-mail: vignolo@dime.unige.it}}
\affiliation{DIME Sez. Metodi e Modelli Matematici, Universit\`{a} di Genova\\
Piazzale Kennedy, Pad.D - 16129 Genova, ITALY}
\date{\today}
\begin{abstract}
We consider the Einstein-Dirac field equations describing a self-gravitating massive neutrino, looking for axially-symmetric exact solutions; in the search of general solutions, we find some that are specific and which have critical features, such as the fact that the space-time curvature turns out to be flat and the spinor field gives rise to a vanishing bi-linear scalar $\overline{\psi}\psi=0$ with non-vanishing bi-linear pseudo-scalar $i\overline{\psi}\gamma^5\psi\not=0$: because in quantum field theory general computational methods are built on plane-wave solutions, for which bi-linear pseudo-scalar vanishes while the bi-linear scalar does not vanish, then the solutions we found cannot be treated with the usual machinery of quantum field theory. This means that for the Einstein-Dirac system there exist admissible solutions which nevertheless cannot be quantized with the common prescriptions; we regard this situation as yet another issue of tension between Einstein gravity and quantum principles. Possible ways to quench this tension can be seen either in enlarging the validity of quantum field theory or by restricting the space of the solutions of the Einstein-Dirac system of field equations.
\end{abstract}
\maketitle
\section{Introduction}
In modern physics, the field equations of many theories arise from an action functional where the Lagrangian embodies all the necessary information on the dynamics of one or more interacting fields. Such amount of information is encoded into the resulting field equations, describing the physical features of the system.

Finding all exact solutions for a given system of field equations has always been a task very difficult to accomplish: for example, solutions for two fields in interaction amount to just the Kerr solution of an electrodynamic field immerse in its own gravitational field, even if this is not entirely a complete solution since the charged mass distribution is thought to be condensed in the central singularity; complete solutions in which both fields and matter are considered have not been found until very recently. In \cite{CFV}, the system of Dirac fields in interaction with their own gravitational field has been considered in the neutral massless case, and all exact solutions have been found: massless Dirac fields may be useful in describing the dynamics of the plasma we have when fermions have not yet acquired mass due to symmetry breaking in the very early universe. However, it is essential to consider also the system of Dirac fields in interaction with their own gravitational field for neutral but massive particles.

Such particles are neutrinos, thus quite elusive; nevertheless, scattering amplitudes involving $Z^{0}$-loops do take into account external legs in which two neutrino fields are present, and therefore having exact solutions in this case would have a considerable impact. In this paper, we will proceed in finding some of these solutions.

We will follow the same procedure of \cite{CFV}, although now due to the complexity of the task we will not find all exact solutions but only some remarkable instances: our achievement may look incomplete, but we think the found solutions are nevertheless worth noticing in view of the tension between existing theories.

Indeed, one of the most important problems in fundamental physics in modern times is the incompatibility between gravity and quantum field theory (QFT); another face of this problem, or to some extent another problem altogether, is also the seemingly absolute separation of the two domains, with the standard model of particle physics looking totally disconnected from the standard model of cosmology: discrepancies go from the fact that gravity is not quantized nor renormalizable, to the fact that the standard model of particle physics does not account for gravitational effects and when it does the Higgs mechanism gives rise to a cosmological constant largely off the cosmological data \cite{w}.

The list of incompatibilities giving rise to tensions between Einstein gravity and QFT is long, but to an attentive scrutiny some of the arguments arise from reasons that may have no general applicability: for example, according to the current wisdom encoded by the Wilson approach, in order for a theory to be renormalizable it must be such that the dynamical term should remain the leading term in the Lagrangian while going to smaller and smaller scales, but this approach contains the assumption that the interaction in question be expressed in terms of particles, an assumption that does not necessarily hold for gravity, not only for the fact that the graviton is still unobserved but also because gravity is not an interaction but the curvature of the space-time; another instance is the fact that gravity has never been quantized, and again here lies the hypothesis that everything in nature must be quantized, which is a hypothesis that starts to be questioned even among quantum field theorists \cite{DYSON:2013jra}.

Of course, this does not mean that all of the incompatibilities are based on arguments that must be rejected as irrelevant but clearly the tension between these two domains is the most stringent if its rationale is given with the fewest additional hypotheses: here we present a source of tension that is based on no additional hypothesis at all.

More specifically, we will see that the Einstein-Dirac system of coupled field equations has exact solutions that do not respect some of the conditions needed for their quantum-field-theoretical treatment.
\section{Notations and field equations}
In the paper we use the notations of \cite{CFV}. In particular, partial derivatives of a given function $f(x^h)$ are indicated by $f_{x^h}:=\frac{\partial f}{\partial x^h}$. Latin and Greek indices run from $1$ to $4$. The metric tensor of the space-time is denoted by $g_{ij}$, while any tetrad field associated with a given metric is denoted by $e^\mu = e^\mu_i\,dx^i$. In this way, one has $\eta_{\mu\nu}=g_{ij}e^i_\mu e^j_\nu$, where $\eta_{\mu\nu}=\mathrm{diag}(-1,-1,-1,1)$ is the Minkowskian metric and $e^i_\mu$ is the inverse matrix of $e^\mu_i$. The relations $e^i_\mu e^\mu_j=\delta^i_j$ and $e^\mu_i e^i_\nu= \delta^\mu_\nu$ are then verified. Dirac matrices are denoted by $\gamma^\mu$ and $\Gamma^i := e^i_\mu\gamma^\mu$ as well as $\gamma^5 :=i\gamma^4\gamma^1\gamma^2\gamma^3$ are defined. Also, Chiral representation of Dirac matrices is adopted, as well as natural units $\hbar=c=8\pi G=1$ are used. 

The spinorial-covariant derivatives of a neutral Dirac field $\psi$ are expressed as 
\begin{equation}\label{defdsm}
D_i\psi=\psi_{x^i}-\Omega_{{i}}\psi
\end{equation}
where the spinorial-connection coefficients $\Omega_i$ are given by
\begin{equation}\label{1.1}
\Omega_i := - \frac{1}{4}g_{jh}\omega_{i\;\;\;\nu}^{\;\;\mu}e_\mu^j e^\nu_k\Gamma^h\Gamma^k.
\end{equation}
In equation \eqref{1.1}, $\omega_{i\;\;\;\nu}^{\;\;\mu}$ are the spin-connection, associated to a linear connection $\Gamma_{ik}^{\;\;\;j}$ through the usual relation
\begin{equation}\label{1.2}
\Gamma_{ij}^{\;\;\;h} = \omega_{i\;\;\;\nu}^{\;\;\mu}e_\mu^h e^\nu_j + e^{h}_{\mu}\partial_{i}e^{\mu}_{j}.
\end{equation}
We will consider a Dirac field coupled to gravity in the Einstein general relativity theory. The Dirac Lagrangian is
\begin{equation}\label{1.1bis}
L_\mathrm{D} =\frac{i}{2}\left( \overline{\psi}\Gamma^iD_i\psi-D_i\overline{\psi}\Gamma^i\psi\right)-m\overline{\psi}\psi
\end{equation}
where $m$ is the mass of the spinor. Field equations are obtained by varying the Einstein--Hilbert plus the Dirac Lagrangian with respect to the metric and the spinor fields: they turn out to be the Einstein equations
\begin{equation}\label{1.3a}
R_{ij} -\frac{1}{2}Rg_{ij}= \frac{i}{4} \left( \overline{\psi}\Gamma_{(i}{D}_{j)}\psi - {D}_{(j}\overline{\psi}\Gamma_{i)}\psi \right)
\end{equation}
and the Dirac equations
\begin{equation}\label{1.3b}
i\Gamma^{h}D_{h}\psi - m\psi=0
\end{equation}
where $R_{ij}$ and $R$ denote the Ricci tensor and the curvature scalar associated with the Levi--Civita connection. The Einstein equations \eqref{1.3a} can be written in the equivalent form
\begin{equation}\label{1.3c}
R_{ij} = \frac{i}{4} \left( \overline{\psi}\Gamma_{(i}{D}_{j)}\psi - {D}_{(j}\overline{\psi}\Gamma_{i)}\psi \right) - \frac{m}{4}\overline{\psi}\psi\/g_{ij}.
\end{equation}
\section{Massive neutrino in Lewis-Papapetrou space-time}
In the following, we search for solutions of the coupled field equations \eqref{1.3a} and \eqref{1.3b} under the assumption that the geometry of space-time is described by the Lewis--Papapetrou metric in the Weyl canonical coordinates, and the Dirac field represents a massive neutrino in its rest reference frame. The metric tensor is then expressed as   
\begin{equation}\label{Lewis}
{ds}^{2}= -B^2 (d\rho^2 +dz^2) -A^2 (-W\,dt + d\varphi)^2 + \frac{\rho^2}{A^2}\,dt^2
\end{equation} 
where all functions $A(\rho,z)$, $B(\rho,z)$ and $W(\rho,z)$ depend on the $\rho$ and $z$ variables only, while the more general spinor field satisfying the given hypothesis of describing a matter distribution in the rest frame results to be of the form  
\begin{eqnarray}\label{HPL2}
\psi=\left(\begin{tabular}{c}
$\Xi\/e^{i\beta}$\\ $\Lambda\/e^{i\left(\beta+\mu\right)}$\\ $\Xi\/e^{i\left(\beta+\phi\right)}$\\ $\Lambda\/e^{i\left(\beta+\phi+\mu\right)}$
\end{tabular}\right)
\end{eqnarray}
where {\it a priori} $\Xi$, $\Lambda$, $\beta$, $\mu$ and $\phi$ are functions on all variables $\rho,z,\varphi,t$. The co--tetrad field associated to \eqref{Lewis} is
\begin{equation}\label{defcoframe}
e^1 = B\,d\rho, \quad e^2= B\,dz, \quad e^3 = A\left(-W\,dt +d\varphi\right), e^4 = \frac{\rho}{A}\,dt
\end{equation}
and its dual is
\begin{equation}\label{defframe}
e_1 = \frac{1}{B}\,\frac{\partial}{\partial \rho}, \quad e_2 = \frac{1}{B}\,\frac{\partial}{\partial z}, \quad e_3 = \frac{1}{A}\,\frac{\partial}{\partial \varphi}, \quad e_4 = \frac{AW}{\rho}\,\frac{\partial}{\partial \varphi} + \frac{A}{\rho}\,\frac{\partial}{\partial t}.
\end{equation}
The momentum and spin four-vectors of the spinor field can be respectively expressed in terms of the frame \eqref{defframe} as
\begin{equation}\label{PV}
P= \overline{\psi}\gamma^\mu\psi\,e_\mu= \left(2\Xi^2 + 2\Lambda^2\right)\,e_4\qquad {\rm and} \qquad V= \overline{\psi}\gamma^\mu\gamma^5\psi\,e_\mu = 4\Xi\Lambda\cos\mu\,e_1 + 4\Xi\Lambda\sin\mu\,e_2 + \left(2\Xi^2 - 2\Lambda^2\right)\,e_3
\end{equation} 
consistently with the requirement that $P$ singles out the rest frame of the massive neutrino: in fact, the vector $P$ is the velocity density, and (\ref{HPL2}) is simply the most general form of the spinor ensuring the spatial part of $P$ vanishes. 

By imposing that the vectors $P$ and $V$ both possess the same symmetries of the metric \eqref{Lewis}, namely $L_{\frac{\partial}{\partial\varphi}}P=L_{\frac{\partial}{\partial t}}P=0$ and $L_{\frac{\partial}{\partial\varphi}}V=L_{\frac{\partial}{\partial t}}V=0$ (where $L$ denotes Lie derivative), as well as by requiring that the spin vector $V$ is orthogonal to the two Killing vectors $\frac{\partial}{\partial\varphi}$ and $\frac{\partial}{\partial t}$, it necessarily follows that all the quantities $\Xi$, $\Lambda$, $\mu$ and $\phi$ have to be functions of the variables $\rho$ and $z$ only, and moreover $\Xi$ and $\Lambda$ have to coincide, namely
\begin{equation}\label{caso2}
\Lambda=\Xi=\Xi(\rho,z), \quad \mu=\mu(\rho,z) \quad {\rm and} \quad \phi=\phi(\rho,z).
\end{equation} 
The check of \eqref{caso2} is straightforward taking the identities $\overline{\psi}\psi=2\left(\Xi^2 + \Lambda^2\right)\cos\phi$, $i\overline{\psi}\gamma^5\psi=-2\left(\Xi^2 + \Lambda^2\right)\sin\phi$ and $-V^2=P^2=\left[\overline{\psi}\psi^2 + \left(i\overline{\psi}\gamma\psi\right)^2\right]$ into account.

In order to perform covariant derivatives of the Dirac field and then to implement the field equations \eqref{1.3a} and \eqref{1.3b}, we have to evaluate the spinorial connection. In this respect, employing the relations \eqref{1.1} and \eqref{1.2} together with the expression of the metric \eqref{Lewis}, it is seen that in our case the coefficients \eqref{1.1} induced by the Levi--Civita connection are
\begin{subequations}
\begin{equation}
\Omega_{\rho}=\frac{A^2W_{\rho}\gamma^3\gamma^4}{4\rho}+\frac{B_{z}\gamma^1\gamma^2}{2B}
\label{oo1}
\end{equation}
\begin{equation}
\Omega_{z}=-\frac{B_{\rho}}{2B}\gamma^1\gamma^2 + \frac{A^2W_{z}}{4\rho}\gamma^3\gamma^4
\label{oo2}
\end{equation}
\begin{equation}
\Omega_{\varphi} = \frac{A^{3}W_{\rho}}{4\rho\/B}\gamma^1\gamma^4 - \frac{A_{\rho}}{2B}\gamma^1\gamma^3 + \frac{A^{3}W_{z}}{4\rho\/B}\gamma^2\gamma^4 - \frac{A_{z}}{2B}\gamma^2\gamma^3
\label{oo3}
\end{equation}
\begin{eqnarray}
\nn
\Omega_{t} = \left(\frac{AW_{\rho}}{4B} + \frac{WA_{\rho}}{2B}\right)\gamma^1\gamma^3 + \left(\frac{AW_{z}}{4B}+\frac{WA_{z}}{2B}\right)\gamma^2\gamma^3\\
+\left(-\frac{\rho\/A_\rho}{2A^2B}+\frac{1}{2AB}-\frac{A^3WW_{\rho}}{4\rho\/B}\right)\gamma^1\gamma^4
+\left(-\frac{\rho\/A_{z}}{2A^2B}-\frac{A^3WW_{z}}{4\rho\/B} \right)\gamma^2\gamma^4.
\label{oo4}
\end{eqnarray}
\label{def01}
\end{subequations}
Now, taking conditions \eqref{caso2} into account, the field equations can be written in full. Dirac equations \eqref{1.3b} evaluated for the metric \eqref{Lewis} and for spinor \eqref{HPL2}, and decomposed in their real and imaginary parts, amount to eight real equations
\begin{subequations}\label{eqdd}
\begin{equation}\label{eqdd1}
\begin{split}
\frac{B_z\Xi}{2B^2} - \frac{W_zA^2\Xi}{4B\rho} + \frac{\Xi_z - \Xi\beta_\rho -\Xi\phi_\rho - \Xi\mu_\rho}{B} - \frac{A\Xi\cos\mu\left(W\beta_\varphi + \beta_t\right)}{\rho} - \frac{\Xi\beta_\varphi\cos\mu}{A} \\
- m\Xi\cos\phi\cos\mu + m\Xi\sin\phi\sin\mu =0 
\end{split} 
\end{equation}
\begin{equation}\label{eqdd2}
\begin{split}
\frac{B_\rho\Xi}{2B^2} - \frac{W_\rho\/A^2\Xi}{4B\rho} + \frac{\Xi_\rho + \Xi\beta_z + \Xi\phi_z + \Xi\mu_z}{B} + \frac{\Xi}{2B\rho} + \frac{A\Xi\sin\mu\left(W\beta_\varphi + \beta_t\right)}{\rho} + \frac{\Xi\beta_\varphi\sin\mu}{A}\\
m\Xi\sin\phi\cos\mu + m\Xi\cos\phi\sin\mu =0 
\end{split}
\end{equation}
\begin{equation}\label{eqdd3}
\begin{split}
- \frac{B_z\Xi}{2B^2} - \frac{\Xi\beta_\rho + \Xi\phi_\rho + \Xi_z}{B} - \frac{A\Xi\cos\mu\left(W\beta_\varphi + \beta_t\right)}{\rho} - \frac{W_zA^2\Xi}{4B\rho} + \frac{\Xi\beta_\varphi\cos\mu}{A}\\
- m\Xi\cos\phi\cos\mu - m\Xi\sin\phi\sin\mu =0
\end{split}
\end{equation}
\begin{equation}\label{eqdd4}
\begin{split}
\frac{B_\rho\Xi}{2B^2} + \frac{W_\rho\/A^2\Xi}{4B\rho} + \frac{\Xi_\rho - \Xi\beta_z - \Xi\phi_z}{B} - \frac{A\Xi\sin\mu\left(W\beta_\varphi + \beta_t\right)}{\rho} + \frac{\Xi}{2B\rho} + \frac{\Xi\beta_\varphi\sin\mu}{A}\\
+ m\Xi\sin\phi\cos\mu - m\Xi\cos\phi\sin\mu =0 
\end{split}
\end{equation}
\begin{equation}\label{eqdd5}
\begin{split}
\frac{\Xi\left(- B_z\cos\mu + B_\rho\sin\mu \right)}{2B^2} + \frac{A^2\Xi\left(W_\rho\sin\mu - W_z\cos\mu\right)}{4B\rho} - \frac{A\Xi\left(W\beta_\varphi +\beta_t\right)}{\rho}+
\frac{\Xi\beta_\varphi}{A} + \frac{\Xi\sin\mu}{2B\rho}\\
+ \frac{-\Xi_z\cos\mu + \Xi\beta_\rho\cos\mu + \Xi\mu_\rho\cos\mu + \Xi_\rho\sin\mu + \Xi\mu_z\sin\mu + \Xi\beta_z\sin\mu}{B} -m\Xi\cos\phi =0
\end{split}
\end{equation}
\begin{equation}\label{eqdd6}
\begin{split}
- \frac{\Xi\left(B_\rho\cos\mu + B_z\sin\mu\right)}{2B^2} - \frac{A^2\Xi\left(W_\rho\cos\mu + W_z\sin\mu\right)}{4B\rho} - \frac{\Xi\cos\mu}{2B\rho} - m\Xi\sin\phi \\
\frac{+ \Xi\beta_\rho\sin\mu + \Xi\mu_\rho\sin\mu - \Xi\mu_z\cos\mu - \Xi\beta_z\cos\mu - \Xi_\rho\cos\mu - \Xi_z\sin\mu}{B} =0
\end{split}
\end{equation}
\begin{equation}\label{eqdd7}
- \frac{A\Xi\cos\mu\left(W\beta_\varphi + \beta_t\right)}{\rho} + \frac{B_z\Xi}{2B^2} + \frac{\Xi\beta_\rho + \Xi_z}{B} - \frac{\Xi\beta_\varphi\cos\mu}{A} - \frac{W_zA^2\Xi}{4B\rho} - m\Xi\cos\phi\cos\mu + m\Xi\sin\phi\sin\mu =0
\end{equation}
\begin{equation}\label{eqdd8}
- \frac{A\Xi\sin\mu\left(W\beta_\varphi + \beta_t\right)}{\rho} - \frac{B_\rho\Xi}{2B^2} + \frac{\Xi\beta_z - \Xi_\rho}{B} - \frac{\Xi}{2B\rho} - \frac{\Xi\beta_\varphi\sin\mu}{A} + \frac{W_\rho\/A^2\Xi}{4B\rho} - m\Xi\sin\phi\cos\mu - m\Xi\cos\phi\sin\mu.
\end{equation}
\end{subequations}
Analogously, the Einstein equations in the form \eqref{1.3c} assume the expression
\begin{subequations}\label{eqcampoEx}
\begin{equation}\label{eqcampoEx1}
\frac{BW_\rho\/A^2\Xi^2\sin\mu}{2\rho} - B\Xi^2\phi_\rho\cos\mu - mB^2\Xi^2\cos\phi + \frac{A^4W_\rho^2}{2\rho^2} - \frac{B_{\rho\rho} + B_{zz}}{B} - \frac{2A^2_\rho}{A^2} + \frac{B_z^2 + B_\rho^2}{B^2} + \frac{2A_\rho}{A\rho} + \frac{B_\rho}{B\rho} =0
\end{equation}
\begin{equation}\label{eqcampoEx2}
\frac{BA^2\Xi^2\left(W_z\sin\mu - W_{\rho}\cos\mu\right)}{4\rho} - \frac{B\Xi^2\left(\phi_z\cos\mu + \phi_\rho\sin\mu\right)}{2} + \frac{A^4W_zW_\rho}{2\rho^2} - \frac{2A_zA_\rho}{A^2} + \frac{A_z}{A\rho} + \frac{B_z}{B\rho} =0
\end{equation}
\begin{equation}\label{eqcampoEx4}
\frac{\rho\Xi^2\left(2\beta_\rho + \mu_\rho + \phi_\rho\right)}{2A}=0
\end{equation}
\begin{equation}\label{eqcampoEx5}
- \frac{BW_zA^2\Xi^2\cos\mu}{2\rho} - B\Xi^2\phi_z\sin\mu + \frac{W_z^2A^4}{2\rho^2} - mB^2\Xi^2\cos\phi - \frac{B_{\rho\rho} + B_{zz}}{B} - \frac{B_\rho}{\rho\/B} + \frac{B_\rho^2 + B_z^2}{B^2} - \frac{2A^2_z}{A^2} =0
\end{equation}
\begin{equation}\label{eqcampoEx7}
\frac{\rho\Xi^2\left(2\beta_z + \mu_z + \phi_z\right)}{2A} =0
\end{equation}
\begin{equation}\label{eqcampoEx8}
- \frac{A^4W_\rho\Xi^2\sin\mu}{2\rho\/B} - mA^2\Xi^2\cos\phi - \frac{A^6\left(W_\rho^2 +W_z^2\right)}{2\rho^2B^2} + \frac{A^2_\rho + A^2_z - AA_{\rho\rho} - AA_{zz}}{B^2} - \frac{AA_\rho}{\rho\/B^2} + \frac{A^4W_z\Xi^2\cos\mu}{2\rho\/B} =0
\end{equation}
\begin{equation}\label{eqcampoEx9}
\begin{split}
\frac{\rho\Xi^2\left(A_\rho\sin\mu - A_z\cos\mu\right)}{AB} + \frac{A^4W\Xi^2\left(W_\rho\sin\mu - W_z\cos\mu\right)}{2\rho\/B} + \frac{WA^6\left(W_\rho^2 +W_z^2\right)}{2\rho^2B^2} + \frac{A^2\left(W_{\rho\rho} + W_{zz}\right)}{2B^2} + \frac{\rho\Xi^2\beta_\varphi}{A} \\
+ \frac{AW\left(A_{\rho\rho} + A_{zz}\right) - W\left(A_\rho^2 + A_z^2\right) + 2A\left(A_\rho\/W_\rho + A_zW_z\right)}{B^2} + \frac{WAA_\rho}{\rho\/B^2} - \frac{A^2W_\rho}{2\rho\/B^2} - \frac{\Xi^2\sin\mu}{2B} + mA^2\Xi^2W\cos\phi =0
\end{split}
\end{equation}
\begin{equation}\label{eqcampoEx10}
\begin{split}
\frac{A^4W^2\Xi^2\left(W_z\cos\mu - W_\rho\sin\mu\right)}{2\rho\/B} - \frac{\rho^2\left(A_{\rho\rho}+ A_{zz}\right)}{A^3B^2} + \frac{2\rho\/W\Xi^2\left(A_z\cos\mu - A_\rho\sin\mu\right)}{AB}\\
 + \frac{A^2WW_\rho - AA_\rho\/W^2}{\rho\/B^2} - \frac{\rho\/A_\rho}{A^3B^2}
- \frac{A^2\left(W_\rho^2 + W_z^2\right)}{2B^2} + \frac{\rho^2\left(A_\rho^2 + A_z^2\right)}{A^4B^2} + \frac{\rho\Xi^2\left(W_z\cos\mu - W_\rho\sin\mu\right)}{2B}\\
 - \frac{A^6W^2\left(W_\rho^2 + W_z^2\right)}{2\rho^2B^2} - \frac{4AW\left(W_\rho\/A_\rho + W_zA_z\right)}{B^2} 
- \frac{2\rho\Xi^2\beta_t}{A} - \frac{A^2W\left(W_{\rho\rho} + W_{zz}\right)}{B^2} +\\ \frac{W^2\left(A^2_\rho + A^2_z\right)}{B^2} - \frac{AW^2\left(A_{\rho\rho} + A_{zz}\right)}{B^2} + \frac{W\Xi^2\sin\mu}{B} + \frac{m\rho^2\Xi^2\cos\phi}{A^2} - mA^2W^2\Xi^2\cos\phi =0.
\end{split}
\end{equation}
\end{subequations}
Note that the Einstein equations $\rho\varphi$ and $z\varphi$ are identically verified.
\section{Searching for solutions}
To begin looking for solutions, from equations \eqref{eqcampoEx4} and \eqref{eqcampoEx7} we derive the relations
\begin{subequations}
\begin{equation}\label{dp1}
\phi_\rho = - 2\beta_\rho - \mu_\rho
\end{equation}
\begin{equation}\label{dp2}
\phi_z = - 2\beta_z - \mu_z.
\end{equation}
\end{subequations}
Deriving \eqref{dp1} and \eqref{dp2} with respect $\varphi$ and $t$ and taking \eqref{caso2}, we obtain $\beta_{\rho\varphi}= \beta_{\rho\/t} = \beta_{z\varphi} = \beta_{zt} =0$ from which
\begin{equation}\label{sbb1}
\beta(\rho,z,\varphi,t) = \overline{\beta}(\rho,z) + \hat{\beta}(\varphi,t).
\end{equation}
Inserting equation \eqref{sbb1} into equations \eqref{dp1} and \eqref{dp2} and omitting an inessential integration constant, we get 
\begin{equation}\label{sv6}
\overline{\beta}(\rho,z) = - \frac{\phi}{2} - \frac{\mu}{2}
\end{equation}
and thus 
\begin{equation}\label{sbb2}
\beta = - \frac{\phi}{2} - \frac{\mu}{2} + \hat{\beta}.
\end{equation}
Inserting equation \eqref{sbb2} into equation \eqref{eqdd1} and deriving with respect to $t$, we have
\begin{equation}\label{sv44}
\hat{\beta}_{tt} = - \frac{\hat{\beta}_{\varphi\/t}\left(A^2W+\rho\right)}{A^2}.
\end{equation}
Substituting equation \eqref{sv44} into the derivative with respect to $t$ of the Dirac equations \eqref{eqdd}, it is easily seen that the function $\hat{\beta}$ has to satisfy $\hat{\beta}_{\varphi\/t}=0$ and thus, in view of \eqref{sv44}, we have
\begin{equation}\label{sv3}
\hat{\beta}(\varphi,t) = - Et + V(\varphi)
\end{equation} 
for some function $V(\varphi)$. Inserting \eqref{sbb2} and \eqref{sv3} into the Dirac equations \eqref{eqdd} and deriving with respect to $\varphi$, we obtain $V_{\varphi\varphi}=0$, telling that $V(\varphi)$ is to be linear. Setting to zero some integration constants, we conclude that
\begin{equation}\label{sbb4}
\beta = - \frac{\phi}{2} - \frac{\mu}{2} - Et.
\end{equation}
Inserting the content of equation \eqref{sbb4} into Dirac equations \eqref{eqdd}, it is immediately seen that equations \eqref{eqdd1} and \eqref{eqdd2} are identical to equations \eqref{eqdd7} and \eqref{eqdd8} respectively. Moreover, we can take $\Xi$ of the form
\begin{equation}\label{nXi1}
\Xi = \frac{\alpha\/e^{\frac{s}{2}}}{\sqrt{B}\rho}
\end{equation}
where $\alpha$ is a constant and $s(\rho,z)$ is a function. After inserting equations \eqref{sbb4} and \eqref{nXi1} into Dirac equations, we can subtract \eqref{eqdd7} from \eqref{eqdd3} and \eqref{eqdd8} from \eqref{eqdd4} and add \eqref{eqdd7} to \eqref{eqdd3} and \eqref{eqdd8} to \eqref{eqdd4} to obtain respectively
\begin{subequations}\label{eqdda2b1}
\begin{equation}
\mu_\rho = 2mB\sin\phi\sin\mu + s_z \ \ \ \ \ \ \ \  \mathrm{and}  \ \ \ \ \ \ \ \ 
\mu_z = -2mB\sin\phi\cos\mu - s_\rho + \frac{1}{\rho}
\end{equation}
\begin{equation}\label{dph}
\phi_\rho = - 2mB\cos\mu\cos\phi + \frac{2EAB\cos\mu}{\rho} - \frac{A^2W_z}{2\rho}\ \ \ \  \mathrm{and} \ \ \ \ 
\phi_z = - 2mB\cos\phi\sin\mu + \frac{2EAB\sin\mu}{\rho} + \frac{A^2W_\rho}{2\rho}.
\end{equation}
\end{subequations} 
It is then a straightforward matter to verify that equations \eqref{eqdd5} and \eqref{eqdd6} are automatically implied by equations \eqref{eqdda2b1} which represent (together with \eqref{caso2}, \eqref{sbb4} and \eqref{nXi1}) all the content of Dirac equations. From \eqref{dph} we also get
\begin{equation}\label{dw}
W_\rho = \frac{4mB\rho\cos\phi\sin\mu}{A^2} - \frac{4EB\sin\mu}{A} + \frac{2\rho\phi_z}{A^2}\ \ \ \  \ \ \ \ \mathrm{and} \ \ \ \ \ \ \ \ 
W_z = - \frac{4mB\rho\cos\mu\cos\phi}{A^2} + \frac{4EB\cos\mu}{A} - \frac{2\rho\phi_\rho}{A^2}.
\end{equation}
Returning to the Einstein equations, we insert equation \eqref{nXi1} into the remaining six equations of \eqref{eqcampoEx} so obtaining 
\begin{subequations}\label{eqcca3z}
\begin{equation}\label{eqcca3z1}
\frac{W_{\rho}A^2\alpha^2e^s\sin\mu}{2\rho^3} - \frac{\alpha^2\/e^s\left(\phi_\rho\cos\mu + mB\cos\phi\right)}{\rho^2} + \frac{A^4W_\rho^2}{2\rho^2} - \frac{B_{\rho\rho} + B_{zz}}{B}  - \frac{2A_\rho^2}{A^2} + \frac{B_\rho^2 + B_z^2}{B^2} + \frac{2A_\rho}{\rho\/A} + \frac{B_\rho}{\rho\/B} =0
\end{equation}
\begin{equation}\label{eqcca3z2}
\left(\frac{W_zA^2\alpha^2}{4\rho^3} - \frac{\alpha^2\phi_\rho}{2\rho^2}\right)\/e^s\sin\mu - \left(\frac{W_\rho\/A^2\alpha^2}{4\rho^3} + \frac{\alpha^2\phi_z}{2\rho^2}\right)e^s\cos\mu + \frac{A^4W_\rho\/W_z}{2\rho^2} - \frac{2A_\rho\/A_z}{A^2} + \frac{A_z}{\rho\/A} + \frac{B_z}{\rho\/B} =0
\end{equation}
\begin{equation}\label{eqcca3z3}
- \frac{W_zA^2\alpha^2e^s\cos\mu}{2\rho^3} + \frac{W_z^2A^4}{2\rho^2} - \frac{\alpha^2e^s\left(\phi_z\sin\mu + mB\cos\phi\right)}{\rho^2} - \frac{B_{\rho\rho} + B_{zz}}{B} + \frac{B_\rho^2 + B_z^2}{B^2} - \frac{B_\rho}{\rho\/B} - \frac{2A^2_z}{A^2} =0
\end{equation}
\begin{equation}\label{eqcca3z4}
\frac{A^4\alpha^2e^s\left(W_z\cos\mu - W_\rho\sin\mu\right)}{2\rho^3\/B^2} - \frac{A^6\left(W_\rho^2 + W_z^2\right)}{2\rho^2\/B^2} - \frac{A\left(A_{\rho\rho} + A_{zz}\right)}{B^2} + \frac{A^2_\rho + A^2_z}{B^2} - \frac{mA^2\alpha^2e^s\cos\phi}{\rho^2B} - \frac{AA_\rho}{\rho\/B^2} =0
\end{equation}
\begin{equation}\label{eqcca3z5}
\begin{split}
\left(\frac{A_\rho\alpha^2}{\rho\/AB^2} + \frac{A^4\alpha^2WW_\rho}{2\rho^3B^2} - \frac{\alpha^2}{2\rho^2\/B^2}\right)e^s\sin\mu - \left(\frac{A^4\alpha^2WW_z}{2\rho^3B^2} + \frac{\alpha^2A_z}{\rho\/AB^2}\right)e^s\cos\mu + \frac{A^6W\left(W_\rho^2 + W_z^2\right)}{2\rho^2B^2} + \frac{WAA_\rho}{\rho\/B^2}\\
+ \frac{mW\alpha^2A^2e^s\cos\phi}{\rho^2B} - \frac{W\left(A_\rho^2 +A_z^2\right)}{B^2} + \frac{AW\left(A_{\rho\rho} + A_{zz}\right)}{B^2} + \frac{2A\left(W_\rho\/A_\rho + W_zA_z\right)}{B^2} + \frac{A^2\left(W_{\rho\rho} + W_{zz}\right)}{2B^2}
- \frac{W_\rho\/A^2}{2\rho\/B^2} =0
\end{split}
\end{equation}
\begin{equation}\label{eqcca3z6}
\begin{split}
\left(- \frac{A^4W^2\alpha^2W_\rho}{2\rho^3B^2} - \frac{2W\alpha^2A_\rho}{\rho\/AB^2} - \frac{\alpha^2W_\rho}{2\rho\/B^2} + \frac{W\alpha^2}{\rho^2B^2}\right)e^s\sin\mu + \left(\frac{2W\alpha^2A_z}{\rho\/AB^2} + \frac{A^4W^2\alpha^2W_z}{2\rho^3B^2} + \frac{\alpha^2W_z}{2\rho\/B^2}\right)e^s\cos\mu \\
\left(- \frac{mA^2W^2\alpha^2\cos\phi}{\rho^2B} - \frac{2E\alpha^2}{\rho\/AB} + \frac{m\alpha^2\cos\phi}{A^2B}\right)e^s + \frac{W^2\left(A_\rho^2 + A_z^2\right)}{B^2} - \frac{A^2\left(W_\rho^2 + W^2_z\right)}{2B^2} + \frac{\rho^2\left(A_\rho^2 + A_z^2\right)}{A^4B^2} \\
- \frac{\rho\/A_\rho + \rho^2\left(A_{\rho\rho} + A_{zz}\right)}{A^3B^2} - \frac{AW^2\left(A_{\rho\rho} + A_{zz}\right)}{B^2} - \frac{A^2W\left(W_{\rho\rho} + W_{zz}\right)}{B^2} - \frac{4AW\left(A_\rho\/W_\rho + A_zW_z\right)}{B^2} \\
- \frac{A^6W^2\left(W_\rho^2 + W^2_z\right)}{2\rho^2B^2} + \frac{A^2WW_\rho - AW^2A_\rho}{\rho\/B^2} =0.
\end{split}
\end{equation}
\end{subequations}
Defining the quantities
\begin{equation}
p := \frac{2A^4W}{A^4W^2 - \rho^2}\ \ \ \ \ \ \ \  \mathrm{and} 
\ \ \ \ \ \ \ r := \frac{A^4}{A^4W^2 - \rho^2}
\end{equation}
from the combination $\eqref{eqcca3z4} + p\/\eqref{eqcca3z5} + r\/\eqref{eqcca3z6}$ we get the equation
\begin{equation}\label{eq456x}
\frac{2A^2\alpha^2e^s\left(-m\rho\cos\phi + EA\right)}{\rho\/B\left(A^4W^2 - \rho^2\right)} =0
\end{equation}
yielding the relation
\begin{equation}\label{scosphi}
\cos\phi = \frac{EA}{m\rho}.
\end{equation}
Making use of \eqref{scosphi}, equations \eqref{eqdda2b1} simplify to 
\begin{subequations}\label{edda3b}
\begin{equation}
\mu_\rho = 2mB\sin\phi\sin\mu + s_z \ \ \ \ \ \ \ \  \mathrm{and} \ \ \ \ \ \ \ \ 
\mu_z = -2mB\sin\phi\cos\mu - s_\rho + \frac{1}{\rho}
\end{equation}
\begin{equation}\label{edda3b3}
\phi_\rho =- \frac{A^2W_z}{2\rho}\ \ \ \ \ \ \ \  \mathrm{and} \ \ \ \ \ \ \ \ 
\phi_z = \frac{A^2W_\rho}{2\rho}.
\end{equation}
\end{subequations}
For later use, we rewrite equations \eqref{eqdda2b1} also in the form
\begin{subequations}\label{eqdda22}
\begin{equation}\label{eqdda22a}
\mu_\rho = \frac{2B\sqrt{m^2\rho^2 - E^2A^2}\sin\mu}{\rho} + s_z 
\ \ \ \ \ \ \ \  \mathrm{and} \ \ \ \ \ \ \ \ 
\mu_z = - \frac{2B\sqrt{m^2\rho^2 - E^2A^2}\cos\mu}{\rho} - s_\rho + \frac{1}{\rho}
\end{equation}
\begin{equation}\label{eqdda22b}
- \frac{EA_\rho}{\sqrt{m^2\rho^2 - E^2A^2}} + \frac{EA}{\rho\sqrt{m^2\rho^2 - E^2A^2}}=- \frac{A^2W_z}{2\rho}\ \ \ \ \ \ \ \  \mathrm{and} \ \ \ \ \ \ \ \ 
- \frac{EA_z}{\sqrt{m^2\rho^2 - E^2A^2}} = \frac{A^2W_\rho}{2\rho}.
\end{equation}
\end{subequations}
Going back to the remaining Einstein equations \eqref{eqcca3z}, multiplying by $\frac{2}{\rho}A^2B^2$ the combination $W$\eqref{eqcca3z4} + \eqref{eqcca3z5}, we get
\begin{equation}\label{ee45}
\frac{4A^3\left(A_\rho\/W_\rho + A_zW_z\right)}{\rho} + \frac{A^4\left(W_{\rho\rho}+W_{zz}\right)}{\rho} - \frac{A^2\alpha^2e^s\sin\mu}{\rho^3} - \frac{A^4W_\rho}{\rho^2} + \frac{2A\alpha^2e^s\left(A_\rho\sin\mu - A_z\cos\mu\right)}{\rho^2} =0.
\end{equation}
Let us now define the quantities
\begin{equation}\label{defv1defv2}
v_1\/(\rho,z):= \frac{A^2\alpha^2e^s\sin\mu}{\rho^2} + \frac{A^4W_\rho}{\rho}\ \ \ \ \ \ \ \  \mathrm{and} \ \ \ \ \ \ \ \ 
v_2(\rho,z):= - \frac{A^2\alpha^2e^s\cos\mu}{\rho^2} + \frac{A^4W_z}{\rho}.
\end{equation}
Making use of equations \eqref{edda3b}, a direct check shows that equation \eqref{ee45} amounts to the identity $\left(v_1\right)_\rho + \left(v_2\right)_z =0$ implying the existence of a function $\Lambda(\rho,z)$ satisfying $v_1 =\Lambda_z$ and $v_2 = - \Lambda_\rho$. From these and \eqref{defv1defv2} we derive
\begin{equation}\label{solW12}
W_\rho = - \frac{\alpha^2e^s\sin\mu}{\rho\/A^2} + \frac{\rho\Lambda_z}{A^4}\ \ \ \ \ \ \ \  \mathrm{and} \ \ \ \ \ \ \ \ 
W_z = \frac{\alpha^2e^s\cos\mu}{\rho\/A^2} - \frac{\rho\Lambda_\rho}{A^4}
\end{equation}
together with
\begin{subequations}\label{solescA}
\begin{equation}\label{solescA1}
e^s = \frac{\rho\left(A^4W_z +\rho\Lambda_\rho\right)}{A^2\alpha^2\cos\mu}
\end{equation}
\begin{equation}\label{solescA2}
\sin\mu = \frac{\cos\mu\left(-A^4W_\rho + \rho\Lambda_z\right)}{A^4W_z + \rho\Lambda_\rho}.
\end{equation}
\end{subequations}
Equations \eqref{solescA} make sense under the condition $\cos\mu\not= 0$: this assumption is not restrictive, because under the condition $\cos\mu = 0$ there are no solutions of the field equations. The proof is long and we omit it for brevity.

We can rewrite the remaining Einstein equations \eqref{eqcca3z}, making use of \eqref{scosphi}, \eqref{edda3b} and \eqref{solescA}, according to
\begin{subequations}\label{eccxx} 
\begin{equation}\label{eccxx1}
\frac{W_\rho\Lambda_z + W_z\Lambda_\rho}{2\rho} + \frac{A^4W_z^2}{2\rho^2} - \frac{BA^3EW_z}{\rho^2\cos\mu} - \frac{BE\Lambda_\rho}{\rho\/A\cos\mu} - \frac{B_{\rho\rho} + B_{zz}}{B} + \frac{B_\rho^2 + B^2_z}{B^2} - \frac{2A^2_\rho}{A^2} + \frac{2A_\rho}{\rho\/A} + \frac{B_\rho}{\rho\/B} =0
\end{equation}
\begin{equation}\label{eccxx2}
- \frac{A^4W_\rho\/W_z}{2\rho^2} + \frac{\left(W_z\Lambda_z - W_\rho\Lambda_\rho\right)}{2\rho} - \frac{2A_\rho\/A_z}{A^2} + \frac{A_z}{\rho\/A} + \frac{B_z}{\rho\/B} =0
\end{equation}
\begin{equation}\label{eccxx3}
- \frac{W_z\Lambda_\rho + W_\rho\Lambda_z}{2\rho} + \frac{A^4W_\rho^2}{2\rho^2} - \frac{BA^3EW_z}{\rho^2\cos\mu} - \frac{BE\Lambda_\rho}{\rho\/A\cos\mu} - \frac{B_{\rho\rho}+B_{zz}}{B} + \frac{B^2_\rho + B^2_z}{B^2} - \frac{B_\rho}{\rho\/B} - \frac{2A^2_z}{A^2} =0
\end{equation}
\begin{equation}\label{eccxx4}
- \frac{A^2W_\rho\Lambda_z}{2\rho\/B^2} - \frac{A^5EW_z}{\rho^2\/B\cos\mu} - \frac{AE\Lambda_\rho}{\rho\/B\cos\mu} - \frac{A\left(A_{\rho\rho}+A_{zz}\right)}{B^2} + \frac{A^2_\rho +A^2_z}{B^2} - \frac{AA_\rho}{\rho\/B^2} + \frac{A^2W_z\Lambda_\rho}{2\rho\/B^2} =0
\end{equation}
\begin{equation}\label{eccxx5}
\begin{split}
\frac{A\left(W_\rho\/A_\rho + W_zA_z\right)}{B^2} + \frac{\rho\left(A_\rho\Lambda_z - A_z\Lambda_\rho\right)}{A^3B^2} + \frac{A^2W\left(W_\rho\Lambda_z - W_z\Lambda_\rho\right)}{2\rho\/B^2} - \frac{\Lambda_z}{2A^2B^2} + \frac{WA^5EW_z}{\rho^2\/B\cos\mu} + \frac{WAE\Lambda_\rho}{\rho\/B\cos\mu} \\
+ \frac{WA\left(A_{\rho\rho} + A_{zz}\right)}{B^2} - \frac{W\left(A_\rho^2 + A^2_z\right)}{B^2} + \frac{A^2\left(W_{\rho\rho}+W_{zz}\right)}{2B^2} + \frac{WAA_\rho}{\rho\/B^2} =0
\end{split}
\end{equation}
\begin{equation}\label{eccxx6}
\begin{split}
\frac{A^2W^2\left(W_z\Lambda_\rho - W_\rho\Lambda_z\right)}{2\rho\/B^2} + \frac{2\rho\/W\left(A_z\Lambda_\rho - A_\rho\Lambda_z\right)}{A^3B^2} - \frac{\rho\/E\Lambda_\rho}{A^3B\cos\mu} - \frac{AW^2E\Lambda_\rho}{\rho\/B\cos\mu} - \frac{A^5W^2EW_z}{\rho^2B\cos\mu} - \frac{AEW_z}{B\cos\mu}\\
- \frac{2AW\left(W_\rho\/A_\rho + W_zA_z\right)}{B^2} + \frac{\rho\left(W_z\Lambda_\rho - W_\rho\Lambda_z\right)}{2A^2B^2} - \frac{AA_\rho\/W^2}{\rho\/B^2} + \frac{W\Lambda_z}{A^2B^2} - \frac{\rho\/A_\rho}{A^3B^2} - \frac{AW^2\left(A_{\rho\rho}+A_{zz}\right)}{B^2} \\
- \frac{A^2W\left(W_{\rho\rho}+W_{zz}\right)}{B^2} - \frac{\rho^2\left(A_{\rho\rho}+A_{zz}\right)}{A^3B^2} + \frac{\rho^2\left(A^2_\rho + A^2_z\right)}{A^4B^2} + \frac{W^2\left(A^2_\rho + A^2_z\right)}{B^2} =0.
\end{split}
\end{equation}
\end{subequations}
At this point of the discussion it is convenient to distinguish the two distinct cases $E\not =0$ and $E=0$.  
\subsection{The case $\boldsymbol{E\neq0}$}
In this case, we can search for solutions for $A$ and $W$ in terms of an unknown function $T(\rho,z)$, in the form
\begin{subequations}
\begin{equation}\label{rtt1}
A = \frac{2\rho\/mE}{\sqrt{m^4\rho^2\/T^2_\rho + 4E^4}}
\end{equation}
\begin{equation}\label{rtt2}
T_z =W.
\end{equation}
\end{subequations}
Indeed, on the one hand expressions \eqref{rtt1} and \eqref{rtt2} can be inserted into \eqref{eqdda22b} yielding that one equation is verified and the other gives the Laplace equation for $T(\rho,z)$
\begin{equation}\label{dt11}
T_{\rho\rho} + T_{zz} + \frac{T_\rho}{\rho} =0.
\end{equation}
The conclusion is that $T(\rho,z)$ has to be harmonic and so, from \eqref{rtt2}, $W(\rho,z)$ must be harmonic too. Moreover, it seems physically meaningful requiring that for large values of $\rho$ and $z$ the metric \eqref{Lewis} has to reduce to Minkowskian space-time, and then, if $W$ is regular anywhere (with respect to the adopted coordinates), we necessarily conclude that $W=0$. Indeed, in such a circumstance equation \eqref{rtt2} implies that $T$ is function of the variable $\rho$ only, namely $T=T(\rho)$. Moreover, going back to equations \eqref{eqdda22b}, we get that $A$ is a linear function of the variable $\rho$ only. Omitting an inessential integration constant, we have $A(\rho) = c_1\rho$. Inserting $W=0$ and $A=c_1\rho$ into equations \eqref{eccxx4}, \eqref{eccxx5} and \eqref{eccxx6}, it is easily seen that the identities $\Lambda_\rho=\Lambda_z =0$ have to hold necessarily. But this is not possible due to equations \eqref{solW12}. The conclusion follows that for $W$ regular anywhere we have no solutions. Alternatively, the function $W$ can be non zero if and only if we give up its full regularity anywhere. This case is currently under study.  
\subsection{The case $\boldsymbol{E=0}$}
In this case, equations \eqref{eqdda22b} yield $W$ constant and then $W=0$. Also, from equation \eqref{scosphi} we have $\phi =\pi/2$ (or $\phi = 3\pi/2$). Moreover, equations \eqref{eqdda22a} reduce to
\begin{equation}\label{eddxx}
\mu_\rho = 2mB\sin\mu + s_z\ \ \ \ \ \ \ \  \mathrm{and} \ \ \ \ \ \ \ \ 
\mu_z = -2mB\cos\mu - s_\rho + \frac{1}{\rho}.
\end{equation}
We can look for functions of the form $A =\rho\/e^{q(\rho,z)}$ and $B = e^{b(\rho,z)}$ where $q(\rho,z)$ and $b(\rho,z)$ are unknown functions, and by employing these definitions, with \eqref{scosphi} and \eqref{eddxx}, as well as $W=0$ and $\phi =\pi/2$, Einstein equations \eqref{eqcca3z} become
\begin{subequations}\label{eccxxx}
\begin{equation}\label{eccxxx1}
- b_{\rho\rho} - b_{zz} -\frac{2q_\rho}{\rho} -2q_\rho^2 + \frac{b_\rho}{\rho} =0
\end{equation}
\begin{equation}\label{eccxxx2}
- \frac{q_z}{\rho} - 2q_\rho\/q_z + \frac{b_z}{\rho} =0
\end{equation}
\begin{equation}\label{eccxxx3}
- b_{\rho\rho} - b_{zz} - \frac{b_\rho}{\rho} - 2q^2_z =0
\end{equation}
\begin{equation}\label{eccxxx4}
- \frac{\rho^2e^{2q(\rho,z)}}{e^{2b(\rho,z)}}\left(q_{\rho\rho} + q_{zz} + \frac{q_\rho}{\rho}\right) =0
\end{equation}
\begin{equation}\label{eccxxx5}
\frac{e^s\alpha^2\sin\mu}{2\rho^2e^{2b(\rho,z)}} + \frac{e^s\alpha^2q_\rho\sin\mu}{\rho\/e^{2b(\rho,z)}} - \frac{e^s\alpha^2q_z\cos\mu}{\rho\/e^{2b(\rho,z)}}=0
\end{equation}
\begin{equation}\label{eccxxx6}
-\frac{1}{e^{2q(\rho,z)}e^{2b(\rho,z)}}\left(q_{\rho\rho} + q_{zz} + \frac{q_\rho}{\rho}\right)=0.
\end{equation}
\end{subequations} 
We notice that \eqref{eccxxx4} and \eqref{eccxxx6} are equivalent and coincide with the Laplace equation for the unknown $q(\rho,z)$
\begin{equation}\label{ecc4}
q_{\rho\rho} + q_{zz} + \frac{q_\rho}{\rho} =0.
\end{equation}
Moreover, directly from  equation \eqref{eccxxx2} we get
\begin{equation}\label{ecc2}
b_z = 2q_\rho\/q_z + q_z.
\end{equation} 
Adding and subtracting equations \eqref{eccxxx1} and \eqref{eccxxx3}, we obtain respectively
\begin{equation}\label{ecc1}
b_\rho = - q^2_z\rho + q_\rho^2\rho + q_\rho
\end{equation}
and
\begin{equation}\label{ecc3}
2b_{\rho\rho} + 2b_{zz} + \frac{2q_\rho}{\rho} + 2q^2_\rho + 2q^2_z =0.
\end{equation}
Making use of equation \eqref{ecc4}, it is a straightforward matter to verify that expressions \eqref{ecc2} and \eqref{ecc1} satisfy the conditions of Schwarz theorem. Equations \eqref{ecc2} and \eqref{ecc1} are then  solvable by quadratures, once the function $q(\rho,z)$ is known. Also, equation \eqref{ecc3} results to be automatically verified if \eqref{ecc4}, \eqref{ecc2} and \eqref{ecc1} are. Finally, from \eqref{eccxxx5} we have
\begin{equation}\label{solmu}
\mu = \arctan\left(\frac{2q_z\rho}{2q_\rho\rho +1}\right).
\end{equation}
The whole problem is now reduced to the solution of Laplace equation \eqref{ecc4}. We discuss some particular case:
\begin{itemize}
\item $q(\rho,z) =c\ln\rho$: For such a choice of function $q(\rho,z)$ we have immediately $A=\rho^{c+1}$, $b=\left(c^2+c\right)\ln\rho$, $\mu=0$ and $s=-\frac{2m\rho^{c^2+c+1}}{c^2+c+1} + \frac{\ln\left(\rho^{c^2+c+1}\right)}{c^2+c+1}$. Consequently, the metric assumes the form
\begin{equation}\label{Lewis3}
ds^2 = -\rho^{2(c^2+c)}\,d\rho^2 - \rho^{2(c^2+c)}\,dz^2 - \rho^{2(c+1)}\,d\varphi^2 + \rho^{-2c}\,dt^2
\end{equation}
and the Dirac field results to be
\begin{eqnarray}\label{Diracsoluition3}
\psi=\left(\begin{tabular}{c}
$\frac{\alpha\/e^{-\frac{2m\rho^{(c^2+c+1)} - \ln\left({\rho^{(c^2+c+1)}}\right)}{2(c^2+c+1)}}}{\sqrt{\rho^{(c^2+c)}}\rho}\/e^{-i\frac{\pi}{4}}$\\ \\
$\frac{\alpha\/e^{-\frac{2m\rho^{(c^2+c+1)} - \ln\left({\rho^{(c^2+c+1)}}\right)}{2(c^2+c+1)}}}{\sqrt{\rho^{(c^2+c)}}\rho}\/e^{-i\frac{\pi}{4}}$\\ \\
$\frac{\alpha\/e^{-\frac{2m\rho^{(c^2+c+1)} - \ln\left({\rho^{(c^2+c+1)}}\right)}{2(c^2+c+1)}}}{\sqrt{\rho^{(c^2+c)}}\rho}\/e^{i\frac{\pi}{4}}$\\ \\
$\frac{\alpha\/e^{-\frac{2m\rho^{(c^2+c+1)} - \ln\left({\rho^{(c^2+c+1)}}\right)}{2(c^2+c+1)}}}{\sqrt{\rho^{(c^2+c)}}\rho}\/e^{i\frac{\pi}{4}}$
\end{tabular}\right)
\end{eqnarray}
and a direct check on the Riemann tensor associated with \eqref{Lewis3} shows that values of the constant $c$ for which we have asymptotic flatness do not exist. For $c=-1$ the Riemann tensor vanishes, so we obtain a space-time diffeomorphic to the Minkowskian one. Finally, for $c=0$ we obtain that $q(\rho,z) =0$ o that $A=\rho$, $\mu =0$, $B=1$ (without loss of generality), $\beta =-\pi/4$, and from equations \eqref{eddxx} we get $s(\rho)=\ln\rho -2m\rho$. So summarizing all the results, for the metric \eqref{Lewis} and the spinor field \eqref{HPL2} we find the solutions respectively given by the Minkowskian metric  in cylindrical coordinates
\begin{equation}\label{Minkowskian}
ds^2 = -\left(d\rho^2 +dz^2\right) - \rho^2\,d\varphi^2 + dt^2
\end{equation}
and the Dirac field of the form
\begin{eqnarray}\label{Diracsoluition}
\psi=\left(\begin{tabular}{c}
$\frac{\alpha\/e^{-m\rho}}{\sqrt{\rho}}\/e^{-i\frac{\pi}{4}}$\\ \\
$\frac{\alpha\/e^{-m\rho}}{\sqrt{\rho}}\/e^{-i\frac{\pi}{4}}$\\ \\
$\frac{\alpha\/e^{-m\rho}}{\sqrt{\rho}}\/e^{i\frac{\pi}{4}}$\\ \\
$\frac{\alpha\/e^{-m\rho}}{\sqrt{\rho}}\/e^{i\frac{\pi}{4}}$
\end{tabular}\right)
\end{eqnarray}
for which constraints given by $\overline{\psi}\psi=0$ but $i\overline{\psi}\gamma^5\psi\not =0$ hold identically.
\item $q(\rho,z) = -\frac{\ln\rho}{2} + c_1z + c_2$: In this case we have $b(\rho,z)= - \frac{c_1^2\rho^2}{2} - \frac{\ln\rho}{4} +c_3$ and $\mu=\frac{\pi}{2}$ and so from \eqref{eddxx} we get $s_\rho = \frac{1}{\rho}$ and $s_z = - \frac{2me^{-\frac{c_1^2\rho^2}{2}+c_3}}{\rho^{\frac{1}{4}}}$ and a direct calculation shows that in this case the Schwarz theorem does not hold and then no solution exist.
\item $q(\rho,z) =\frac{1}{\sqrt{\rho^2 + z^2}}$: In such a circumstance, we have that from equations \eqref{ecc2}, \eqref{ecc1} and \eqref{solmu} we deduce that $b(\rho,z) = \frac{2\rho^4 + 4\rho^2z^2 + 2z^4 - \sqrt{\rho^2 +z^2}\rho^2}{2\left(\rho^2 +z^2\right)^{\frac{5}{2}}}$ and $\mu(\rho,z) = - \arctan\left(\frac{2\rho\/z}{\left(\rho^2 + z^2\right)^{\frac{3}{2}} - 2\rho^2}\right) $ and inserting them into \eqref{eddxx}, we get two equations for the unknown $s(\rho,z)$ and again this does not satisfy integrability conditions.
\item $q(\rho,z) =\frac{cz}{\left(\rho^2 +z^2\right)^{\frac{3}{2}}}$: The situation is identical to the last two cases, so given the explicit expression of the function $q(\rho,z)$, we derive that of $b(\rho,z)$ and those of the first derivatives $\mu_\rho$ and $\mu_z$ of the function of $\mu(\rho,z)$ and inserting all the results into equations \eqref{eddxx}, we get the non-integrability of the equations.
\end{itemize}
\section{Comments on the compatibility with QFT}
In obtaining the previous results, we have found a special class of solutions, corresponding to the choice $E\!=\!0$ and $q(\rho,z) =c\ln\rho$ giving rise to specific solutions which correspond to the Minkowskian metric in cylindrical coordinates and the Dirac field of the form (\ref{Diracsoluition}) so that the curvature tensor vanishes and both scalar and pseudo-scalar bi-linear spinorial quantities vanish as well: what this means is that in looking for exact solutions of spinors in presence of a curved space-time, we have stumbled upon spinors for which the space-time is flat and therefore for these spinors it becomes possible to talk about plane-waves verifying the additional $\overline{\psi}\psi=0$ and $i\overline{\psi}\gamma^5\psi\not =0$ constraints. This situation gives rise to a discussion on a possible interface between the present results and QFT, as we are going to show next.

In QFT, the most important quantity that we should be able to calculate is the scattering amplitude for a given process, because from it all cross sections and decay rates can be evaluated 
\cite{p-s}; the scattering amplitude of a process is obtained as a perturbative expansion given in terms of loop diagrams in which all information about the interaction is condensed in each vertex while every propagator is taken for free fields (for a general discussion about QFT and the calculation of scattering amplitudes see specifically chapter $4$ of \cite{p-s}); the fact that propagators are taken for free fields means that they are the Green functions of the free field equations and thus the fields can be taken in the form of plane wave solutions. As a prototypical example, in the calculation of the magnetic moment of the electron, we have that the scattering amplitude is given in terms of the following expression
\begin{eqnarray}
\mathcal{M}\!\approx\!
qF_{1}(k^{2})A^{\mathrm{ext}}_{\mu}(k)\overline{u}(p')\boldsymbol{\gamma}^{\mu}u(p)
+\frac{q}{2m}F_{2}(k^{2})\frac{1}{2}F^{\mathrm{ext}}_{\mu\nu}(k)
\frac{i}{2}\overline{u}(p')[\boldsymbol{\gamma}^{\mu},\boldsymbol{\gamma}^{\nu}]u(p)
\end{eqnarray}
where $F_{1}(k^{2})$ and $F_{2}(k^{2})$ are the form factors as functions of the square momentum transfer $k=p'-p$ while $A^{\mathrm{ext}}_{\mu}(k)$ is the external electrodynamic potential with momentum $k$ and $\overline{u}(p')$ and $u(p)$ are the initial and final electron states of momenta $p'$ and $p$ calculated in momentum space; in this expression the spinor fields are taken in the form of plane wave solutions of the free Dirac equation after a Fourier transformation. In the form of plane wave solutions, however, one obtains that $i\overline{u}\gamma^5u=0$ identically and that it is always possible to choose $\overline{u}u=2m$ as normalization condition assumed to calculate the completeness relationships that will be needed for computing the sum over all possible spin states (once again we refer to chapter $3$ of \cite{p-s}). Therefore in QFT the common paradigm of calculation dramatically uses the assumption of plane-waves such that $i\overline{u}\gamma^5u=0$ and with $\overline{u}u=2m$ as normalization. This is precisely the issue we have been anticipating above about the interface between QFT and the solutions we have found here.

In order to give QFT a perturbative theory the condition of dealing with plane wave solutions is necessary and such a condition is automatically verified by the solutions we found, and in this sense we can claim that our solution can naturally be treated within the perturbative methods of QFT; on the other hand, in QFT details of the calculations for many of the most important processes involve additional constraints such as $i\overline{u}\gamma^5u=0$ and $\overline{u}u=2m$ which cannot be compatible with the constraints $\overline{\psi}\psi=0$ and $i\overline{\psi}\gamma^5\psi\not =0$ of the solutions we have found here: this means that there are fields that cannot enter into the domain of applicability of QFT as we know it, and yet they are exact solutions of an accepted system of field equations. As a consequence, we regard this as a critical situation lying at the interface between QFT in its standard formulation and the Einsteinian theory of gravity coupled to the Dirac field.

Are there ways out of this situation? Of course, one of the possibility would be to cope with the fact that QFT in its usual structure is powerful enough to cover most of the observed phenomenology but at the same time not general enough to include cases as those we found here: to this purpose it could be possible to wonder whether there exists a version of QFT that could perform calculations also for fields featured by the constraints $i\overline{u}\gamma^5u\neq0$ and $\overline{u}u=0$ or, more in general, without the need of developing in plane-waves. A complementary way out could be to keep QFT as it is, but finding a way so that solutions like those we found above do not appear: to restrict the possible solutions it is necessary to endow with more dynamical terms the Einstein-Dirac system of field equations.

Clearly, this tension may also be resolved by a combination of these two proposals.
\section{Conclusion}
In the present work, we have considered the system of Einstein-Dirac equations looking for exact solutions, in axially-symmetric space-times. We have found some of them, which have the property of giving rise to a flat space-time and a non-trivial spinor field such that $\overline{\psi}\psi=0$ but $i\overline{\psi}\gamma^5\psi\not =0$: the vanishing of the Riemann tensor was not imposed as a reasonable approximation but derived to be an exact result, and therefore we have demonstrated that despite everything interacts with its own gravity nevertheless there can exist matter which turns out to be free; this free matter field results to be a type-$3$ spinor in Lounesto classification \cite{L}, certainly non-standard although not singular \cite{VFC,daRocha:2013qhu}. Matter fields are treated in QFT by finding solutions in the form of plane-waves: for them, the pseudo-scalar vanishes and the normalization condition $\overline{\psi}\psi=2m$ is typically assumed. Instead, here we have found a flat space-time, in which certainly plane-waves can exist, but where there are massive spinors for which $\overline{\psi}\psi=0$ identically and thus the normalization condition $\overline{\psi}\psi=2m\neq0$ cannot hold. The solution we have found does not have at least some of the features that are necessary to be treated in QFT; yet, they are solutions of the field equations of gravity coupled to matter: the solutions we found are admissible but nevertheless they cannot be treated quantum-field-theoretically at least in the form of QFT usually considered. This implies that the common prescriptions of QFT are not of total generality, which may be seen as yet another point of tension between quantization and gravity.

In order for this tension to be smoothed one then can consider the following two ways out: one possibility is to accept that QFT is not general enough to treat all types of particles; the other is to keep QFT as it is, and then go on in restricting the Einstein-Dirac field equations in some suitable way. For instance, this can be done by assuming that the Lagrangian of gravity be different from the Hilbert one: there is quite a list of theories of this type, from general modified gravity \cite{Clifton:2011jh, Capozziello:2011et, Carroll:2004de}, to $f(R)$-gravity \cite{Sotiriou:2008rp, Capozziello:2009nq, Olmo:2011uz}, non-minimally coupled models \cite{Amendola:1999qq, Uzan:1999ch}, conformal theories \cite{Mannheim:2011ds}; in some situations, some of these theories may also solve the problem of gravitational renormalizability \cite{s}, although introducing unitarity violation and ghosts. If we retain the hypothesis of least-order derivative, the other way out is to consider torsion, and because the partially-conserved axial-current gives a coupling between the divergence of torsion and $i\overline{\psi}\gamma^5\psi$, then torsion-spin interactions affect the dynamics of spinors, such dynamics will be restricted and consequently some solutions like those we found here may well be precluded.

It is remarkable how a simple specific solution may give us such an amount of information about physics: it is one more instance telling us that either our understanding of QFT is circumstantial and it should be generalized or that the field equations of the model should be empowered, for example by modifying the Hilbert Lagrangian or more appropriately by not neglecting the torsion of the space-time.

It is clearly impossible to speculate at the moment which one of these alternative scenarios is to be followed although it is most likely it will be a combination of all.

\end{document}